# Acoustic-optic Q-switched cavityless weak-feedback laser based on Nd:GdVO$_4$ bounce geometry

Rui Guo,[1] Qiang Liu,[1,2] Mali Gong[1,3,*]

[1] State Key Laboratory of Precision Measurement and Instruments, Tsinghua University, Beijing 100084
[2] Key Laboratory of Photon Measurement and Control Technology, Ministry of Education, Tsinghua University, Beijing 100084, China
[3] State Key Laboratory of Tribology, Tsinghua University, Beijing 100084
*Corresponding author: gongml@mail.tsinghua.edu.cn

An AOQ (acoustic-optic Q-switched) laser with cavityless weak-feedback configuration is demonstrated based on a Nd:GdVO$_4$ bounce geometry. The laser can operate at a repetition up to 500 kHz with the output power above 4W. The pulse-width at 100 kHz reaches 5.2 ns, which is 4.3 times the round-trip time. A theory of Q-switched cavityless weak-feedback laser is proposed for the first time, to our best knowledge. This theory is very suitable for analyzing short pulses comparable to round-trip time. By utilizing the theory, simulation is implemented for our experimental conditions. The consistence between the simulation and the experimental results proves the validity of our theory.

## 1. Introduction

Lasers are ordinarily generated by adopting a pair of cavity mirrors, which can provide the positive feedbacks. However, under enough high gain, the spontaneous emission can still grow to useful ASE output without cavity[1-3]. This kind of source is also termed as mirrorless laser[4], superfluorescent source[5], or cavityless laser[6]. Since in 1966, study on neodymium-glass laser system with cavityless configuration was reported[1], but limited to the media of low gain level in the past, the system was relatively complex and huge. In the past decades, many types of high gain solid-state lasers has developed rapidly, such as innoslab lasers[7-9], multi-pass laser amplifiers[10-12] and bounce geometry[13-15] lasers. These high gain laser modules make it possible to simplify the cavityless laser system. In 2006, G. Smith et al. demonstrated an ASE source by adopting a high gain Nd:YVO$_4$ bounce geometry[16, 17]. Without an output mirror, continuous wave (CW) mode output of several watts could be directly generated with the near diffraction-limit beam quality. In 2018, Xiaoming Chen et al. presented a narrow-linewidth ASE by adopting multi-pass zigzag Nd:YAG amplifier configuration[18]. By pumping the gain medium with QCW (quasi-continuous-wave) mode, the output can operate at repetition of 100 Hz. However, up to date, rare studies about such cavityless configuration were investigated for Q-switch mode lasers.

In this paper, experimental and theoretical studies on a Q-switched cavityless laser with weak feedbacks are implemented. Experimentally, an AOQ (acoustic-optics Q-switched) cavityless laser with weak feedbacks is demonstrated based on a Nd:GdVO$_4$ bounce geometry. Under a high gain level, pulsed laser of hundreds of kHz repetition can be realized without output cavity mirror. Only weak feedbacks of backscattering caused by the exiting surface of the laser crystal can build up the laser. In an effective optical path length of 180 mm, we realize a pulsed laser with a minimum pulse-width of 5.2 ns, which is 4.3 times the corresponding round-trip time.

Theoretically, in order to describe the characteristics of short pulses produced by cavityless weak-feedback laser, a theory suitable for analyzing Q-switched cavityless weak-feedback lasers is proposed. In traditional Q-switched laser theory, the rate equations for inverted population and photon density are analyzed on a minimum time unit of one round-trip[19]. However, when the pulse-widths have reached only several round-trips, like in our AOQ cavityless laser with weak-feedback, it is obvious that the traditional Q-switched rate equation theory is not suitable to solve the problem of the short Q-switched pulses. By adopting our theory, the resolution of pulse profile will not be limited to the round-trip time. The theory will be very helpful for short pulses analyzing and designing in many high gain laser systems.

## 2. Experimental Setup

Our experimental setup is illustrated in Fig. 1. The cavityless weak-feedback laser basically adopts the structure of a bounce geometry oscillator[14], but no output coupler is needed in this setup.

After the beam exits from the crystal, it enters into an optical isolator (OI) in order to prevent the feedbacks from latter optical surfaces. The output power is monitored by a power meter. The pulse profile is monitored by a photodetector, which receives the signal scattered by the power meter. The beam quality is measured by a M$^2$ beam quality analyzer. The spectrum is analyzed by optical spectrum analyzer and Fabry-Perot interferometer.

## 3. Theoretical analysis and numerical simulation

In the traditional Q-switched pulse theory, the laser operation is described by rate equations of inverted population and photon density[19, 20]:

$$\frac{dN}{dt} = R_p(N_t - N) - \frac{N}{\tau} - W_{up} \cdot N^2 \qquad (1)$$

$$\frac{d\varphi}{dt} = \frac{L_{gain}}{L_{cav}} \cdot N\sigma v\varphi - \frac{\varphi}{\tau_c}. \qquad (2)$$

Here, $N$ and $\varphi$ is the inverted population density and the photon density, respectively. $R_p$ is the pump rate. $N_t$ is the total population density of the gain medium. $\sigma$ is the stimulated emission cross-section. $\tau$ is the lifetime of the upper energy level. $W_{up}$ is the ETU (energy transfer upconversion) coefficient. $v$ is the light speed in the gain medium. $L_{gain}$ is the physical length of the gain medium. $L_{cav}$ is the effective optical path of the cavity. $\tau_c$ is the lifetime of the cavity, which is given by:

$$\tau_c = \frac{t_r}{ln(1/R) + L}, \qquad (3)$$

where $t_r$ is the round-trip time of the cavity, and $L$ is the round-trip parasitic loss in the cavity. $R$ is the reflectivity of the output mirror.

In traditional Q-switched pulse theory, Eq. (1)-(3) are derived at a minimum time scale of one round-trip time (see appendix A in [19]). Thus, the formation of a pulse is analyzed on a minimum time resolution of one round-trip time. In most Q-switched lasers, the pulse-widths are several tens of round-trip time, and this theory is very effective. However, in many high gain laser systems, such as cavityless configuration laser in this work, the pulse-widths are capable of reaching only a few round-trips. It is obvious that a theory suitable for analyzing short Q-switched pulses comparable to round-trip time should be investigated.

In order to characterize the evolution of output intensity and inverted population for short pulses, the time resolution should be less than round-trip time. Based on this requirement, the process of light growth, loss and modulation of Q-switch is concerned within one round-trip time, and a theory suitable for Q-switched cavityless weak-feedback lasers is proposed.

### A. Theoretical analysis of cavityless weak-feedback lasers

A diagram of simplified cavityless laser with weak feedbacks is illustrated in Fig. 2. Assume that the dimension of pump area is $L_{gain} \times W_{gain} \times H_{gain}$ in the laser crystal. $L_{gain}, W_{gain}$ and $H_{gain}$ corresponds to the dimension of length, width and thick, respectively. When the laser crystal is under a high gain condition, the ASE effect is obvious. $I_{ASE}(t)$ represents the intensity of single-pass ASE at the end face of the crystal, which can be obtained by the Linford formula[21]:

$$I_{ASE}(t) = I_s \frac{\Omega}{4} \cdot \frac{G_0(t)}{(ln(G_0(t)))^{1/2}}. \qquad (4)$$

Here $I_s = \frac{hv}{\sigma\tau}$ represents the saturated intensity of the crystal. $hv$ is the single photon energy of laser wavelength. $\Omega$ corresponds to the solid angle of the ASE in the crystal. $G_0(t) = e^{\sigma \cdot N(t) \cdot L_{gain}}$ is the small-signal gain, where $N(t)$ is the time-dependent average inverted population density of the gain area.

The laser output for a cavityless configuration originates from this ASE effect. In order to further analyze the process of growth, modulation and the transmission for the ASE in one round-trip, several time delays during the round-trip are defined first. Here, $t_{d1}$ represents the time delay between the left side of the gain area and the central position in the Q-switch device. $t_{d2}$ corresponds to the time delay between the central position in the Q-switch device and the reflector M₀. $t_d$ is the round-trip time between M₀ and the left side of the gain area, and $t_r$ represents the round-trip time between M₀ and the right surface (namely the exiting surface) of the crystal. Assume $L_{cav}$ is the optical path between M₀ and exiting surface of the crystal. Using the three lengths $\Delta L_1, \Delta L_2$, and $L_Q$ marked in Fig. 2, $t_{d1}, t_{d2}, t_d$ and $t_r$ can be expressed as follows:

$$t_{d1} = \frac{\Delta L_1 + n_Q \cdot \frac{L_Q}{2}}{c}, t_{d2} = \frac{\Delta L_1 + 2\Delta L_2 + n_Q \cdot L_Q}{c},$$

$$t_d = \frac{2(\Delta L_1 + \Delta L_2 + n_Q \cdot L_Q)}{c}, t_r = \frac{2 \cdot L_{cav}}{c} \qquad (5)$$

Here, $c$ is the speed of light in vaccum. $n_Q$ is the refractive index of the Q-switch device.

Once the single-pass ASE emits from the left side of the gain area, the it will be attenuated by transmission loss and modulated by Q-switch device. The transmission loss efficiency of a round-trip between M₀ and the left side of the gain area is assumed to be $\eta$. Assume the Q-switch is the amplitude modulation type device, and the transmittance function of the Q-switch is $T(t)$. The single-pass ASE will experience once round-trip transmission loss and twice modulation of the Q-switch device before entering the gain area again. After the attenuation and the modulation, a portion of single-pass ASE is ready to entering the gain area again. This portion of ASE should be considered as the seed light of the following amplification process. After the amplification of the seed light, the laser exits at the right surface of the crystal. At this moment, weak feedbacks will be brought by the exiting surface. The feedbacks will affect the seed light in the next round-trip. Thus, a term of weak feedbacks should be included into the seed light, besides the term of initial ASE. In this way, combining the time delays mentioned above, the intensity of the seed light could be written as:

$$I_{seed}(t + t_d) = \eta \cdot T(t + t_{d1}) \cdot T(t + t_{d2}) \cdot$$
$$[I_{ASE}(t) + I_{FB}(t)], \qquad (6)$$

where $I_{FB}(t)$ is the term of weak feedbacks, it represents the intensity at the left side of the gain area brought by feedbacks. In cavityless weak-feedback lasers, the feedbacks are very weak since there is no output coupler. The feedbacks could be introduced by reflections from optical surfaces with anti-reflection films coated, or backscattering from the exiting surfaces.

When the seed light of ASE enters into gain medium again, after a time interval of $\Delta t_{gain} = \frac{n \cdot L_{gain}}{c}$, $I_{seed}(t)$ will be amplified out exponentially, and the output intensity can be written as:

$$I_{out}(t + \Delta t_{gain}) = I_{seed}(t) \cdot G_0(t) \qquad (7)$$

Assume reflectivity caused by the exiting surface is $R$. The feedback intensity $I_{FB}(t)$ can be obtained by considering the reflection at exiting surface and the single-pass amplification:

$$I_{FB}(t + \Delta t_{gain}) = R \cdot I_{out}(t) \cdot G_0(t) \qquad (8)$$

Thinking of Eq. (4), (6) and (7), the rate equation of average inverted population density $N(t)$ can be given as follows:

$$\frac{dN(t)}{dt} = R_p \cdot (N_t - N(t)) - \frac{N(t)}{\tau} - W_{up} \cdot N(t)^2$$
$$- \frac{2 \cdot I_{ASE}(t)}{L_{gain} \cdot hv} - \frac{I_{seed}(t) \cdot [G_0(t) - 1]}{L_{gain} \cdot hv} \qquad (9)$$

In the formula above:
$R_p$ can be written as:

$$R_p = \frac{P_p \cdot \sigma_{abs}}{L_{gain} \cdot H_{gain}} \cdot \frac{1 - exp(-\sigma_{abs} \cdot N_t \cdot W_{gain})}{hv} \qquad (10)$$

where $P_p$ is the pump power and $\sigma_{abs}$ is the absorption cross-section.

Different from the rate equation for low gain Q-switched laser, the rate equation here considers the effect of single-pass ASE. Moreover,

since the single-pass ASE is reflected back into the gain area and will further consume the gain, this portion of consumed inverted population is also considered. The photon density equation in traditional rate equation is replaced by Eq. (4), (6) and (7). In this way, the output intensity can be analyzed within the interval of one round-trip time.

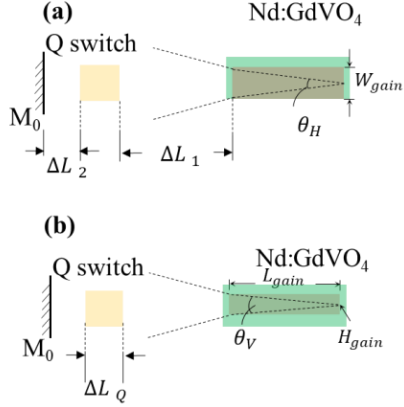

Fig. 2 The simplified model of cavityless weak-feedback laser configuration

## B. Numerical simulations

Combining Eq. (4), (6), (7) and (9), a simulation for pulse characteristics is calculated in this section by adopting parameters in our experiment. Since the bounce angle is less than 5° in our experiment, the grazing gain path is approximated to be straight, like in Fig. 2.

Firstly, a specific expression of transmission loss $\eta$ should be derived. The single-pass ASE occupies a solid angle of $\Omega$, which can be considered on horizontal and vertical directions respectively:

$$\Omega = \theta_H \cdot \theta_V, \quad (11)$$

The divergent angles on horizontal $\theta_H$ and vertical directions $\theta_V$ are determined by the corresponding dimensions of pump area:

$$\theta_H = \frac{W_{gain}}{L_{gain}}, \quad \theta_V = \frac{H_{gain}}{L_{gain}} \quad (12)$$

Assume the cavityless weak-feedback laser is operating on fundamental transverse mode. Thus, only a portion of single-pass ASE corresponds to solid angle of fundamental modes will be not attenuated. Therefore, the transmission efficiency of one round-trip can be written as:

$$\eta = \frac{\theta_{fH} \cdot \theta_{fV}}{\Omega} \quad (13)$$

where $\theta_{fH}$ and $\theta_{fV}$ represent the divergent angles of fundamental mode on horizontal and vertical directions, respectively.

The transmittance function of Q-switch $T(t)$ can be treated as a piecewise function, corresponding to the high loss state and low loss state. When the Q-switch is on high loss state, the transmittance is treated as a constant $T_h$, which characterizes the turn-off capability of a Q-switch device. After the Q-switch opens, the transmittance function is assumed to decrease exponentially. Thus the $T(t)$ can be written as:

$$T(t) = \begin{cases} T_h, & t \in [0, t_h] \\ 1 - (1 - T_h) \cdot e^{-(\frac{t-t_h}{t_{AO}})^2}, & t \in [t_h, T_p] \end{cases} \quad (14)$$

Here, $T_p$ is the repetition period of the modulated signal, and $t_h$ is the duration of high loss state. $t_{AO}$ characterizes the switching time of the AOQ. If $t \to t_h$, the transmittance function $T(t) \to T_h$, and if $t \gg t_h$, $e^{-(\frac{t-t_h}{t_{AO}})^2} \to 0$ thus $T(t)$ approaches to 1.

Since the exiting surface is inclined to the output laser path in our experiment, thus the reflectivity $R$ is mainly caused by the backscattering from the exiting surface of the crystal. It should be noted that even the exiting surface is not perpendicular to the laser path, weak feedbacks caused by diffusions on exiting surface is unavoidable. The reflectivity is determined by the solid angle $\Omega$ and the backscattering reflectivity in unit solid angle $r_\Omega$, which is:

$$R = r_\Omega \cdot \Omega \quad (15)$$

With the knowledge of $\eta$ and $T(t)$, Eq. (4),(6),(7),(9) can be solved numerically.

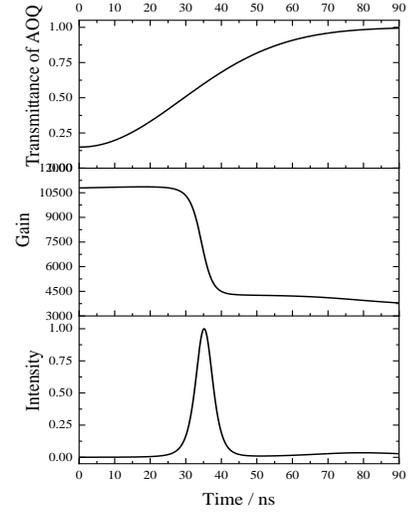

Fig. 3 simulation results in the AOQ cavityless weak-feedback laser for 200 kHz repetition: (a) transmittance of the AOQ; (b) evolution of gain; (c)evolution of a pulse

Fig. 3 shows the simulation results of the AOQ cavityless weak-feedback laser. The repetition rate here is 200 kHz. Here $t$=0 represents the switching time of the AOQ device. Before $t$=0, the transmittance of the Q-switch is 0.15, which is mainly caused by the light undiffracted by the AOQ device. The transmittance rise exponentially after the AOQ switches, as is shown in Fig. 3 (a). Fig. 3 (b) and (c) illustrate the evolution of gain and pulse intensity, respectively. At initial time $t$=0, the gain has been accumulated to $1.05 \times 10^4$. After the Q-switch opens, the gain is released, but very slowly at initial stage. The pulse intensity also grows slowly at this stage. At a critical time (at about $t$=30 ns), the gain overcomes the dropping loss. The gain falls rapidly, and the pulse intensity grows dramatically. The pulse intensity stops growing at a critical point that the gain and loss balance. After that, the output falls to the ground level. The pulse is formed before transmittance of AOQ reaches 100%. The pulse-width (Full Width at Half Maximum) here is 5.5 ns, which is 4.6 times the round-trip time between $M_0$ and exiting surface of Nd:GdVO$_4$. The resolution here is $\Delta t_{gain}$=0.12 ns, which is much less than the round-trip time 1.2 ns. The method here provides a more precise calculation of pulse profiles for short pulses comparable to only a few round-trips.

## 4. Experimental results and discussions

A. Output power and beam quality

The output power of continuous mode versus the pump power is illustrated in the Fig. 4. The threshold pump power is 18 W. The output power rises as with the increase of the pump power. At the pump power of 30 W, the output power reaches 4.5 W, corresponding to the optical-optical efficiency of 15%.

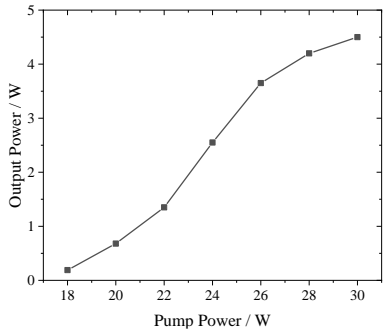

Fig. 4 output power versus pump power (CW mode)

The beam quality under the highest output power is measured by a $M^2$ beam quality analyzer(Spiricon $M^2$-200). The result is shown in the Fig. 5. The $M^2$ factors in horizontal and vertical direction are $M_x^2 < 1.3$ and $M_y^2 < 1.2$, respectively. In the pulsed mode experiment, the beam quality is also monitored and has no obvious variations. The far-field spot profile is shown in the inset. Although there is no output coupler to form a cavity in this experiment, the laser still has good spatial characteristic of near diffraction-limit.

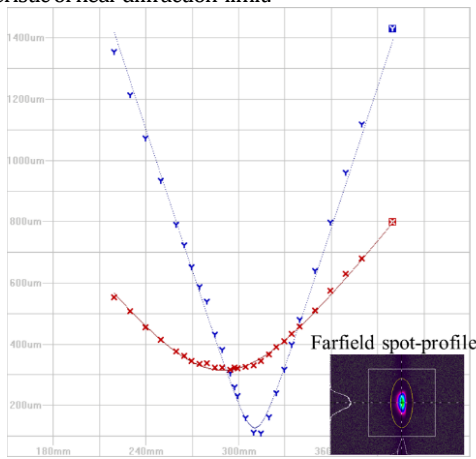

Fig. 5 the beam quality at the output power of 4.5 W

B. Pulse characteristics

The pulsed mode experiment is implemented in this section under the 30 W pump power. Pulse characteristics for repetition rates of 100 – 500 kHz are interested. The releasing time of the AOQ device is 300 ns for all the repetition rates. The output power and the pulse energy is shown in the Fig. 6. It can be found that the output power rises as the increase of the repetition rate. With the repetition rate varying from 100 – 500 kHz, the output power increases from 3.32 W to 4.3 W. The output power rises rapidly before 300 kHz, and after that approaches to saturated power of 4.3 W, which is very close to the output power of CW mode. Under the high loss state, the cavityless weak-feedback laser still has an extinction power of 0.3W. Substracting this portion of extinction power, the pulse energy can be calculated using the output power and repetition rate, which is illustrated as red line in Fig. 6. The pulse energy reaches 30 μJ at 100 kHz, and drops monotonously to 7.5 μJ at the repetition of 500 kHz.

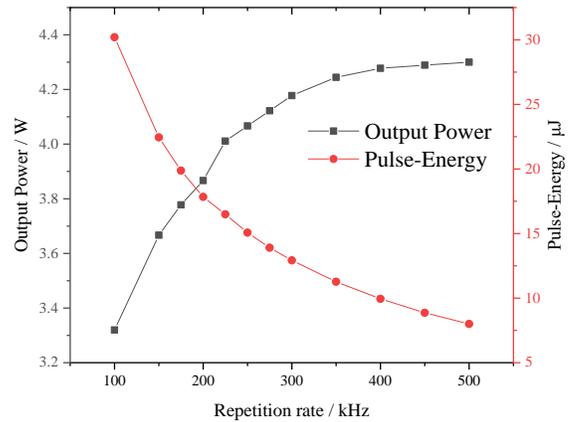

Fig. 6. Output power and Pulse-energy under repetition rates from 100 to 500 kHz

The pulse profiles for 100 – 500 kHz are given in Fig. 7. The experimental waveforms are plotted as solid lines. All the pulse waveforms in our experiment are observed by an InGaAs detector (bandwith of 5 GHz) and an oscilloscope (Tektronix MDO3104, bandwidth of 1 GHz). The pulse profiles are Gaussian shape. There is slight modulations of sub-pulses spacing with $t_r = 1.2$ ns, which is introduced by the weak feedbacks of backscattering from exiting surface. With the repetition rate ranging from 500 kHz to 100 kHz, the pulse intensity becomes stronger, and the pulse-width becomes narrower. This is because the gain is rising as with the increase of accumulating time. For the repetition of 100 kHz, the waveform emerges a subsidiary pulse, which is caused by both impacts of the high gain level and slow switching time of AOQ. The simulated results are ploted as dotted lines in Fig. 7. For 200 – 500 kHz, the simulated profiles fit the measured profiles very well, while for 100 kHz, the position of subsidiary pulse has a deviation from the measured one. We think this deviation might be caused by the deviation of transmittance function of Q-switch between the assumption and the practical situation. Besides the sub-pulse, the profile of main pulse still fits the measured profile well.

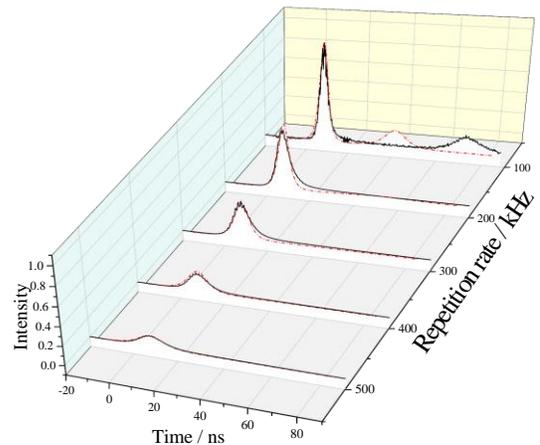

Fig. 7 The pulse profile under repetition rates from 100 to 500 kHz

The pulse-widths for different repetition rates are shown in Fig. 8. The pulse-width rises from 5.2 ns to 13.7 ns monotonously as with the increase of the repetition rate from 100 to 500 kHz, which is illustrated as square markers. The pulse profile is shown in Fig. 8. The simulated results of pulse-widths (calculated in section 3.B) are also plotted in Fig. 8, shown as circle markers. It can be found that the simulated results fit the experimental results well, which proves our theory proposed in section 3.A is effective.

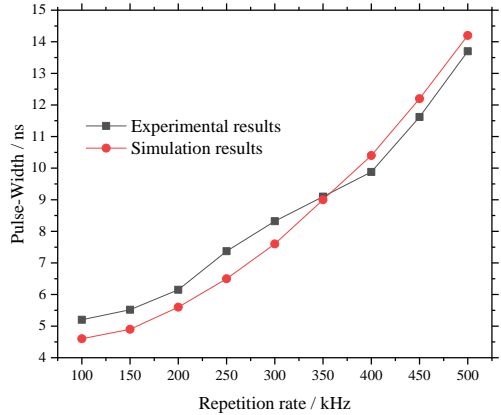

Fig. 8 Experimental and simulated pulse-widths under different repetition rates from 100-500 kHz.

C. Spectral characteristics

The optical spectrum analyzer (Agilent 86142B, minimum resolution of 0.06nm) is applied to measure the spectral characteristics. At the pump power of 30 W, the spectrum of CW mode is recorded. The 10 dB spectral linewidth of the output is measured to be 0.2 nm. The spectral profile is shown in Fig. 9.

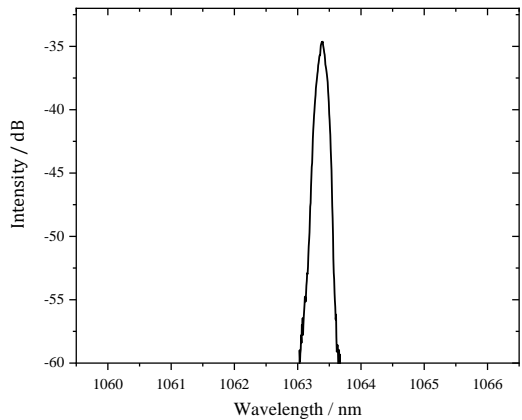

Fig. 9 Spectrum of CW mode cavityless weak-feedback laser output

As the resolution of optical spectrum analyzer is only 0.06 nm, a FP interferometer with free spectral range of 3 GHz is adopted to further investigate the spectral characteristics. When the cavityless weak-feedback laser operates on CW mode (still at the pump power of 30 W), the scanning output of FP interferometer is illustrated as Single frame 1-3 in Fig. 10. It can be found that the spectrum has several disordered peaks: both the amplitudes and the spacings of these peaks are random. No steady longitudinal modes or longitudinal intervals are discovered. However, when an output coupler (T = 80%) is inserted to form a cavity, several steady peaks with the fixed interval are observed in the scanning output. The intervals are consistent with the cavity length between the output coupler and $M_0$. These contrast results prove the random emerging peaks in cavityless weak-feedback laser output are not caused by multiple longitudinal modes. When the scanning output is averaged (such as 64 frames-averaged in Fig. 10), the spectrum becomes flatter. This implies the spectral characteristics in cavityless weak-feedback laser approach to that of ASE in time-average observation.

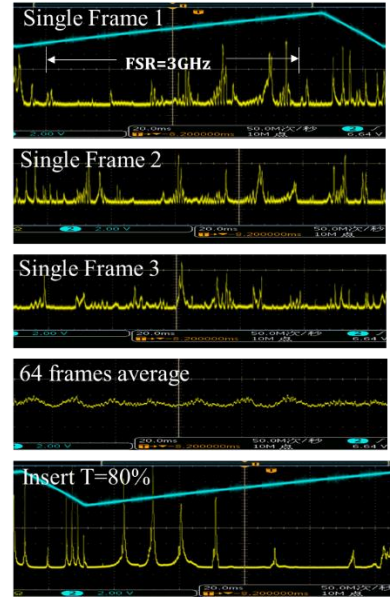

Fig. 10 FP interferometer scanning output of CW mode cavityless weak-feedback laser

## 5. Conclusion

In summary, experimental and theoretical works for an AOQ cavityless weak-feedback laser system are implemented in this paper:

Experimentally, the AOQ cavityless weak-feedback laser system can reach an output power above 4W with the repetition rate up to 500 kHz, owing to the high gain provided by the bounce geometry. The laser has a beam quality of near diffraction limit, and a continuous spectrum in time-averaged observation. The pulse characteristics are also investigated: at 100 kHz, the pulse-width can reach 5.2 ns with the pulse energy of 30 μJ.

On theoretical aspect, a theory for analyzing output characteristics of cavityless weak-feedback laser is proposed for the first time, to our best knowledge. Comparing with traditional rate equations, the theory here can provide a higher resolution of pulse-profile description rather than limited to one round-trip time. By utilizing the theory we proposed, numerical simulations are implemented for our experiment parameters. The numerical simulations are consistent with our experimental results, which proves the validity of our theory.

**Funding Information.** This research is supported by National Key Research and Development Program of China (2017YFB1104500).